\documentclass{pasj01}
\draft 
\Received{$\langle$2022.02.18$\rangle$}
\Accepted{$\langle$2022.08.22$\rangle$}
\Published{$\langle$publication date$\rangle$}
\SetRunningHead{Astronomical Society of Japan}{Usage of \texttt{pasj00.cls}}
\usepackage{bm}
\usepackage{comment}
\usepackage{lineno}
\usepackage{color}
\usepackage{url}

\begin{document}

\title{Polarization calibration of the Tandem Etalon Magnetograph of the Solar Magnetic Activity Research Telescope at Hida Observatory}
\author{Daiki \textsc{Yamasaki},\altaffilmark{1,}$^{*}$ Shin'ichi \textsc{Nagata},\altaffilmark{2} and Kiyoshi \textsc{Ichimoto}\altaffilmark{2}}
\altaffiltext{1}{Astronomical Observatory, Kyoto University, Kitashirakawaoiwake-cho, Kyoto 606-8502, Japan}
\altaffiltext{2}{Hida Observatory, Kyoto University, Kurabashira, Kamitakara-cho, Takayama, Gifu 506-1314, Japan}
\email{dyamasaki@kusastro.kyoto-u.ac.jp}


\maketitle

\begin{abstract}
  The Tandem Etalon Magnetograph (TEM) is one of the instruments of the Solar Magnetic Activity Research Telescope of Hida Observatory.
  The TEM is a partial disk ($320$” $\times240$”) filter magnetograph which scans the wavelength around a Fe I line at $6303$ $\mathrm{\AA}$ and achieves polarimetric sensitivity of $< 5\times10^{-4}$ for each wavelength.
  To obtain the polarimeter response matrix of the instrument, we have carried out end-to-end polarization calibrations of the instrument.
  We have also measured the polarization characteristics of the polarization beam splitter (PBS), which is a crucial component of the instrument.
  As a result of end-to-end calibration, we found significant spatial variation in the response matrix across the field of view.
  From a laboratory test, we found that $1\%$ of the magnitude of a circular diattenuation of the PBS was due to the retardation caused by the stress in the cube and the linear diattenuation of the film. 
  Although the spatial variation across the field of view is more than ten times larger, to achieve the polarimetric sensitivity of $< 5\times10^{-4}$, this can be well explained by the polarization characteristic of the PBS and corrected by using the response matrix obtained in the end-to-end calibration. 
  In addition, we also obtained the daily variation of the polarization property of the TEM.
  We found that the crosstalk from Stokes Q to V changes an amount comparable to the tolerance through a day.
  In the present configuration, we require a pixel-by-pixel calibration every $100$ minutes to meet the accuracy requirement.
\end{abstract}

\section{Introduction}
Solar flares are a rapid magnetic energy release phenomena in the solar corona.
Magnetic reconnection plays an important role in the release of magnetic free energy \citep{Priest2002,Shibata2011}.
During solar flares, some of the accumulated magnetic energy is converted into kinetic energy of the erupting plasmas \citep{Coppi1971,Spicer1982}. 
These erupting plasmas are often observed in the H$\alpha$ line as filament eruptions \citep{Seki2017,Seki2019} and as coronal mass ejections (CMEs: \cite{Parenti2014}) in coronagraphs.
Solar filaments are cool and dense plasma supported by helical coronal magnetic field structures \citep{Xu2012,Hanaoka2017}: magnetic flux ropes (MFRs).
Thus, in order to comprehensively understand the onset mechanisms of solar flares and filament eruptions, we need to clarify the process by which the coronal magnetic field or MFRs become unstable.
We also have to investigate the triggering structures.
Since the coronal magnetic field cannot be directly observed due to observational limitations, the photospheric vector magnetic field observation is crucially important.

According to researches based on the three-dimensional coronal magnetic field extrapolated from the photospheric magnetic field, it is widely believed that magnetohydrodynamical instabilities such as kink instability \citep{Toeroek2004}, torus instability \citep{Kliem2006}, and double-arc instability \citep{Ishiguro2017} play an important role in eruptions of MFRs \citep{Inoue2018,Kusano2020,Yamasaki2021}. 
Theoretical as well as observational studies \citep{Kusano2012,Bamba2013} suggest that the triggering mechanism of some flares is the emergence of a small scale magnetic flux with opposite sign to the magnetic helicity with respect to that of the global magnetic field around the polarity inversion line.
Recent studies on the onset of flares and the triggering and driving mechanisms of MFR eruptions perform the extrapolation of the three-dimensional coronal magnetic field \citep{Bamba2017,Kawabata2017,Woods2018,Muhamad2018,Kang2019,Kawabata2020,Yamasaki2021,Inoue2021}.
In these studies, extrapolations of coronal magnetic field use the photospheric vector magnetic field data taken with either the Solar Optical Telescope (SOT: \cite{Tsuneta2008}; \cite{Suematsu2008}; \cite{Shimizu2008}; \cite{Ichimoto2008}) onboard $Hinode$ \citep{Kosugi2007} or the Helioseismic and Magnetic Imager (HMI: \cite{Schou2012}) onboard the $Solar~Dynamics~Observatory$ ($SDO$: \cite{Pesnell2012}).
However, for both of the instruments, it takes more than $10$ $\mathrm{minutes}$ to obtain the full Stokes parameters with polarimetric sensitivity of $> 10^{-3}$ given the typical spatial size of active regions of $\sim 200\times 200$ $\mathrm{Mm^{2}}$.
On the other hand, in order to investigate the dynamical change of the photostpheric magnetic field caused by a topological change in the corona during the flares, such as the back reaction to the photospheric magnetic field \citep{Hudson2008}, observation with a time cadence of less than 10 minutes is needed.
We also note that the Alfv\'{e}n transit time across the typical coronal loop with a size of $\sim 70$ $\mathrm{Mm}$ is $\sim 1$ $\mathrm{minute}$.
Thus, magnetic field measurements with a wide ($> 200$ $\mathrm{Mm}$) field of view and high temporal cadence of $\sim 1$ $\mathrm{minute}$ are required to investigate the dynamical evolution of the coronal magnetic field with regards to flares and filament eruptions.
Since the change in Lorentz force balance in the solar atmosphere can also be estimated from the photospheric magnetic fields \citep{Fisher2012}, it is expected that we can clearly understand the
triggering structures and the energy release process of solar flares from the observations of the photospheric vector magnetic field with such high temporal cadence.

The Tandem Etalon Magnetograph (TEM: \cite{Nagata2014}) of the Solar Magnetic Activity Research Telescope (SMART: \cite{UeNo2004}) aims to study, with a time cadence of up to 1 minute and a polarimetric sensitivity of $5\times10^{-4}$, the photospheric magnetic field and the magnetic field structure of solar filaments during solar flares.
The TEM obtains the photospheric magnetic field via polarization of the Zeeman effect \citep{delToroIniesta2003, Stenflo2003, Stenflo2013} in the Fe I ($6302.5$ $\mathrm{\AA}$) absorption line. 

Polarization calibration is a method of modeling the polarization of the telescope and the other optical system included in the instrument. 
Polarization calibration is required for precise polarimetric measurements in solar observation \citep{Ichimoto2008,Ishikawa2014,Anan2018,Harrington2019}.
In some cases, they use the instrument calibration unit placed on a focal plane to calibrate the optical system other than the telescope \citep{Iglesias2016}, and in some cases, end-to-end calibrations are performed by placing a polarizer in front of a telescope.
In end-to-end calibration, we examine a polarimeter response matrix \citep{Elmore1990}.
  The advantage of this method is that a polarimeter response matrix combines the stack of Mueller matrices of optical components of the instrument, and also includes the matrix representing the polarimeter’s modulation and demodulation.
  Thus, it is a convenient single matrix type characterization of the measurement system as a whole.
In our research, we performed an extensive polarization calibration of the whole instrument in order to achieve the polarimetric accuracy that meets the TEM's sensitivity.

In this paper, we describe the methodology used for calibrating the TEM polarization and report the final polarization characteristics of the TEM. 
The rest of this paper is structured as follows: the instrument is introduced in Section 2, the experiments are described in Section 3, data analysis and results are presented in Section 4, and discussions arising from our findings are summarized in Section 5.

\section{Instrument}
\subsection{Overview of the polarimeter}
\begin{figure}[h]
  \begin{center}
    \includegraphics[bb= 0 0 1120 470, width=130mm]{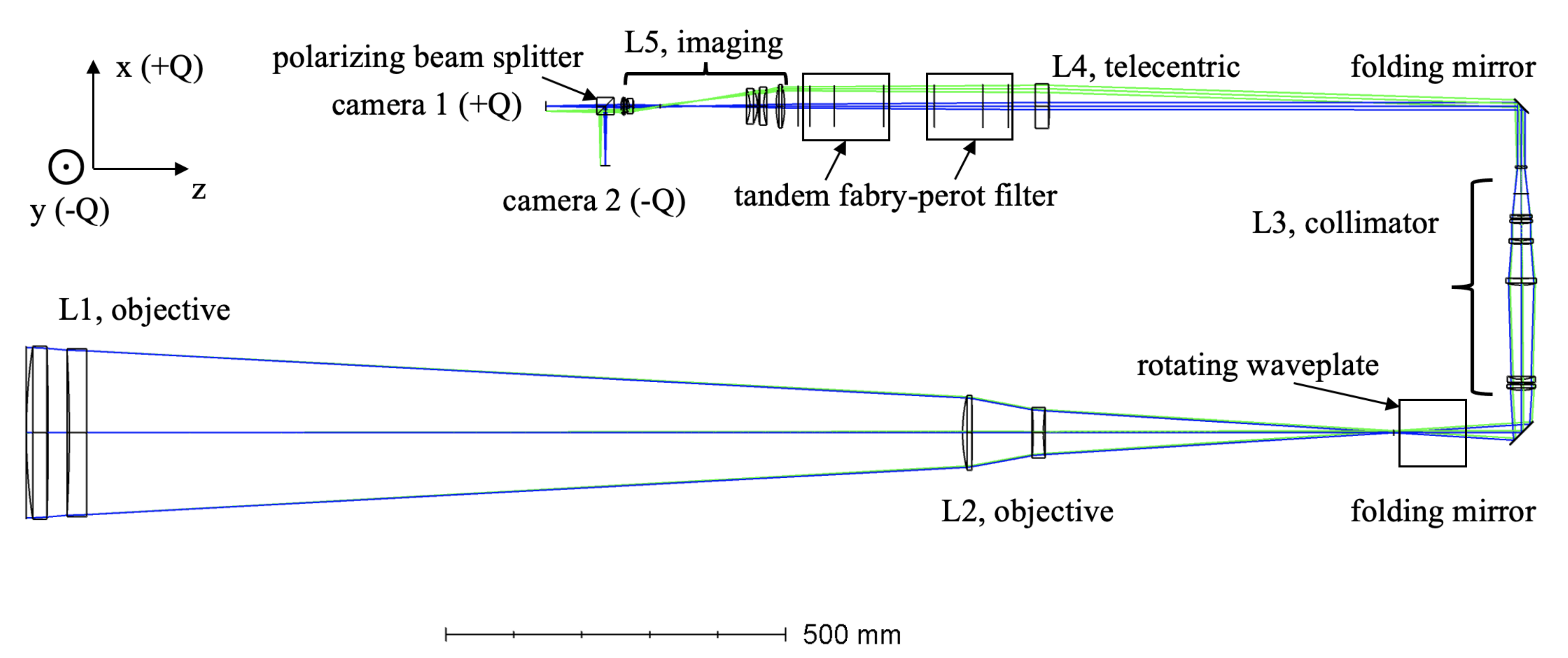}
  \end{center}
  \caption{Optical layout of the TEM. Blue and green lines corespond to the ray at the center and the edge of the field of view, respectively. z-axis: incident light direction, x-axis: $+Q$, y-axis: $-Q$.}\label{fig:OL}
\end{figure}

\begin{figure}[h]
    \begin{center}
      \includegraphics[bb= 0 0 1400 400, width=120mm]{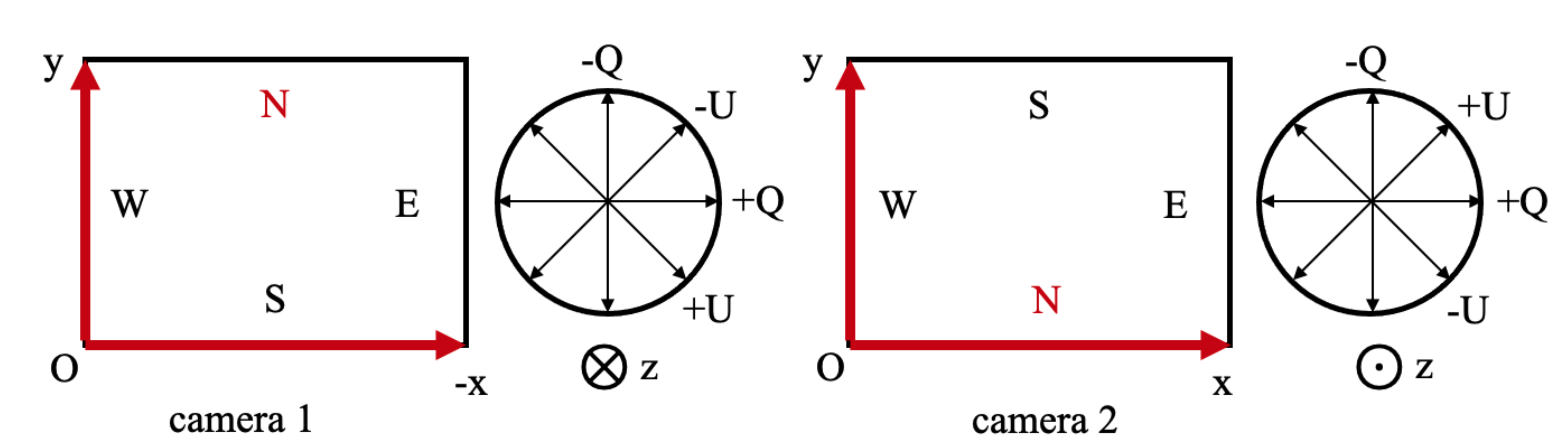}
      \end{center}
    \caption{Definition of coordinate system on two CCD cameras of TEM.}\label{fig:Xdef}
\end{figure}

SMART of Hida Observatory consists of four telescopes equipped to a unique equatorial mount.
The TEM, the fourth telescope, is a $250$ $\mathrm{mm}$ diameter telescope equipped with a magnetograph using tandem Fabry-Perot filters.
In figure \ref{fig:OL}, we show the optical layout of the TEM.
The polarization modulator of the TEM is a rotating waveplate placed just behind the primary focus.
The waveplate is continuously rotating at a frequency of $1.7$ $\mathrm{Hz}$ during observations.
The waveplate is a $94.28$ $\mathrm{\mu m}$-thick crystal attached to a $2.058$ $\mathrm{mm}$-thick S-BSL7 substrate.
The retardation of the waveplate at the working wavelength of $6302.5$ $\mathrm{\AA}$ (Fe I) is designed to be $127^{\circ}$ at $25$ $^{\circ}\mathrm{C}$.
In order to meet the accuracy requirement of the TEM, the retardation should be in the range of $127\pm0.38$ $^{\circ}$.
We note that the ambient temperature of TEM is set at $25$ $^{\circ}\mathrm{C}$ and is controlled by air in $25\pm5^{\circ}\mathrm{C}$.
Thus, we can meet the accuracy requirement (see also \cite{Nagata2014}). 
We take monochromatic images around $6302.5$ $\mathrm{\AA}$ with a tandem Fabry-Perot filter (the bandwidth is $\sim130$ $\mathrm{m\AA}$).
Since the Fabry-Perot filters are made of z-cut lithium niobate and are used in telecentric configuration, these filters do not change polarization state significantly. 
The polarization analyzer of the TEM is a polarizing beam splitter (PBS: Melles Griot 03 PBB 005), with which we divide the beam into orthogonally polarized rays of light; we use the coordinate system that defines $+Q$ as the direction of $p$ component of the PBS in this study (figure \ref{fig:Xdef}).
The PBS is made of a multi-layer film with substrate of BK7 and it has $0.01$ of an extinction ratio in wavelength range of $4500-6800$ $\mathrm{\AA}$. 
Note that the PBS is located on the image side of the telecentric optical system. 
Two CCD cameras placed behind the PBS simultaneously take each of the orthogonally polarized lights with a frame rate of 30 frames per second.

\subsection{Measurement of Stokes vector} \label{sec:intro2}
\begin{figure}[h]
    \begin{center}
      \includegraphics[bb= 0 0 1200 900, width=130mm]{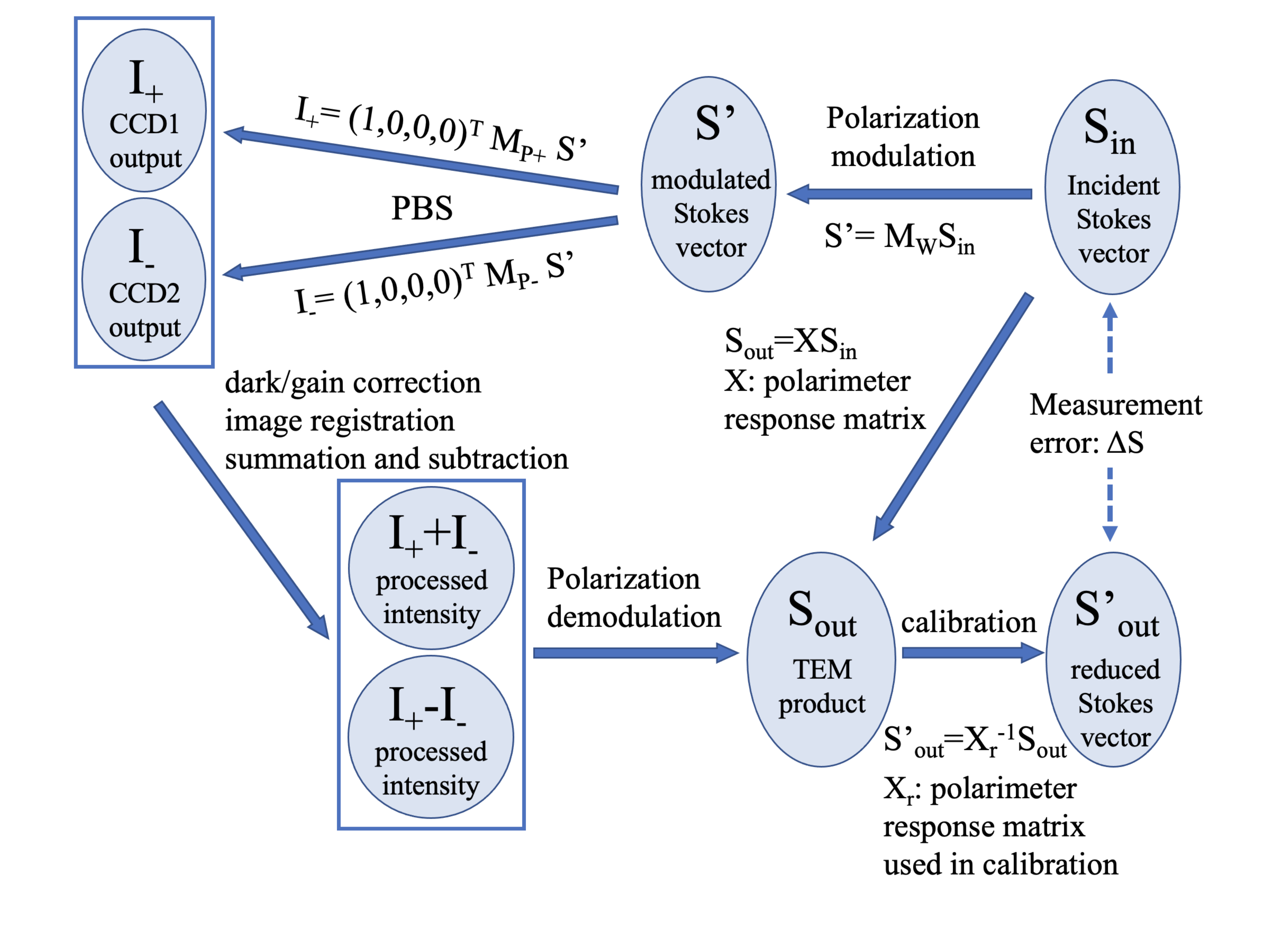}
    \end{center}  
    \caption{Definition of polarimeter response matrix and error in polarization measurement.}\label{fig:CM}
\end{figure}

The procedure to obtain the Stokes vectors with the TEM is explained in figure \ref{fig:CM}.
We derive the Stokes vector ${\bf S_{\mathrm{out}}}$ based on the simple theoretical model as follows.
The observed intensities from the two cameras ($I_{\pm}$) at each wavelength can be given as the following:
\begin{eqnarray}
     I_{\pm}(\theta) &=& {\bf E^{I}} {\bf M^{p\pm}} {\bf M^{w}}(\theta) {\bf S^{\mathrm{in}}} \label{eq:MM} \\
                    &=& \frac{1}{2}\{I_{\mathrm{in}}\pm\left(\frac{1+\cos\delta}{2}+\frac{1-\cos\delta}{2}\cos4\theta\right)\times Q_{\mathrm{in}}\nonumber \\
                    &\pm&\frac{1-\cos\delta}{2}\sin4\theta\times U_{\mathrm{in}}\mp\sin\delta\sin2\theta\times V_{\mathrm{in} }\}, \label{eq:MM2}
\end{eqnarray}
where ${\bf E^{I}}=(1,0,0,0)^{\mathrm{T}}$, ${\bf S^{\mathrm{in}}}=(I_{\mathrm{in}},Q_{\mathrm{in}},U_{\mathrm{in}},V_{\mathrm{in}})^{\mathrm{T}}$ is the Stokes vector at the entrance of the polarization modulator, and ${\bf M^{p\pm}}$ are the Mueller matrices of the PBS, with $\pm$ representing each of the dual orthogonal beams.
${\bf M^{w}(\theta)}$ is the Mueller matrix of the rotating waveplate at the rotating angle of $\theta(t)=\omega t$, where $\omega$ is the angular frequency of the rotating waveplate.
$\delta$ is the retardation of the waveplate ($\delta\simeq127^{\circ}.0$ around $6302.5$ $\mathrm{\AA}$).
The waveplate with this retardation shows the same modulation coefficient for both linear and circular polarization.

Thus, the summation and subtraction of these intensities at each wavelength can be described by the following equations:
\begin{eqnarray}
    I_{+}(\theta)+I_{-}(\theta) &=& I_{\mathrm{in}}, \label{eq:sum}\\
    I_{+}(\theta)-I_{-}(\theta) &=& \frac{1-\cos\delta}{2}Q_{\mathrm{in}}\times\cos4\theta \nonumber \\
    &+&\frac{1-\cos\delta}{2}U_{\mathrm{in}}\times\sin4\theta \nonumber \\
    &-&\sin\delta V_{\mathrm{in}}\times\sin2\theta. \label{eq:sub}
\end{eqnarray}
Here we note that registration among the dual-beam images is performed by using 3 parameters of $x-$direction shift, $y-$direction shift, and rotation angle which are determined by an experiment in sub-pixel accuracy.

The advantage of the two orthogonal polarized beam measurements is that we can separate the time series of $I_{\mathrm{in}}$ and $Q_{\mathrm{in}}/U_{\mathrm{in}}/V_{\mathrm{in}}$ by using a summation and a subtraction of two components.
Thus, we can suppress the crosstalk from the incident Stokes $I$ to the others.
Since Stokes $I$ is larger than the others, the suppression of the crosstalk from Stokes $I$ due to image jitter during data accumulation is a crucial technique to derive accurate magnetic field measurements.

Based on the theoretical model given by equations (\ref{eq:sum}) and (\ref{eq:sub}), we derive the output Stokes vector ${\bf S_{out}}$ by fitting a sinusoidal function of $\theta$, $2\theta$, and $4\theta$ which correspond to angular frequency of $\omega, 2\omega$ and $4\omega$.
Here we perform demodulation after combining two orthogonally polarized beams since we need to make registration of images to suppress the guiding error during the data acquisition by using the combined and unmodulated intensity map.

However, the actual Mueller matrix of the TEM might deviate from the theoretical expression given by equation (\ref{eq:MM}).
Therefore, we need to calibrate ${\bf S_{out}}$ with the polarimeter response matrix ${\bf X}$\citep{Elmore1990}.
The response matrix ${\bf X}$ is defined as follows in order to represent the crosstalk between incident and output Stokes parameters.
\begin{eqnarray}
  {\bf S_{in}} ={\bf X^{-1}}{\bf S_{out}}.
\end{eqnarray}
The purpose of polarization calibration is to examine the relationship between the known incident Stokes vector and the corresponding output Stokes vector.

\section{Experiments}
\subsection{On the accuracy and sensitivity of polarization measurement}
\citet{Ichimoto2008} discuss the accuracy requirement of the measurement of the polarimeter response matrix.
The errors of the measured normalized Stokes parameter $S/I$ $(=1, Q/I, U/I,$ and $V/I)$ can be expressed by
\begin{equation}
    \Delta \left(\frac{S}{I}\right)=\Delta_{\mathrm{s}}\left(\frac{S}{I}\right)+\Delta_{\mathrm{b}}, \label{eq:error}
\end{equation}
where the first term in the right hand side of the equation is the error in scale, and the second term is the error in bias.
Here we note that the error in polarization calibration is the error of the measured polarimeter response matrix ${\bf X}$.
Thus the error due to the measured response matrix ${\bf X_r}$ can be described as
\begin{eqnarray}
    \Delta\left(\frac{S}{I}\right)=\left({\bf X_{r}^{-1}}{\bf X}-{\bf E}\right)\left(\frac{S}{I}\right), \nonumber
\end{eqnarray}
where ${\bf X}$ is the true polarimeter response matrix, ${\bf X_{r}}$ is the measured one, and ${\bf E}$ is the identity matrix.

We require that $\Delta_{\mathrm{s}}$ be smaller than the given allowable scale error $a$, and $\Delta_{\mathrm{b}}$ be smaller than the expected photometric noise $\varepsilon$.
Thus, the response matrix is required to follow the inequality, 
\begin{equation}
    |{\bf X_{r}^{-1}}{\bf X}-{\bf E}| = |{\bf \Delta} {\bf X}| < 
    \left(
    \begin{array}{cccc}
        -     & a/P_{\mathrm{l}} & a/P_{\mathrm{l}} & a/P_{\mathrm{c}} \\
        \varepsilon & a & \varepsilon/P_{\mathrm{l}} & \varepsilon/P_{\mathrm{c}} \\
        \varepsilon & \varepsilon/P_{\mathrm{l}} & a & \varepsilon/P_{\mathrm{c}} \\
        \varepsilon & \varepsilon/P_{\mathrm{l}} & \varepsilon/P_{\mathrm{l}} & a
    \end{array}
    \right), \nonumber
\end{equation}
with regards to measurement error. 
$P_{\mathrm{l}}$ and $P_{\mathrm{c}}$ are the expected maximum linear- and circular-polarization degrees in the solar spectra, respectively \citep{Anan2018}.

In the case of TEM's polarization calibration, we adopt $\varepsilon=5.0\times10^{-4}, a=0.05, P_{\mathrm{l}}=0.15,$ and $P_{\mathrm{c}}=0.20$ for the accuracy requirement, and the tolerance matrix of ${\bf X}$ is described below.
\begin{equation}
\label{eq:req}
    |{\bf \Delta} {\bf X}| < 
    \left(
    \begin{array}{cccc}
        -      & 0.3333 & 0.3333 & 0.2500 \\
        0.0005 & 0.0500 & 0.0035 & 0.0025 \\
        0.0005 & 0.0035 & 0.0500 & 0.0025 \\
        0.0005 & 0.0035 & 0.0035 & 0.0500
    \end{array}
    \right).
\end{equation}
This requirement is higher than that of Hinode/SOT \citep{Ichimoto2008}, because our photometric accuracy of $\varepsilon=5.0\times10^{-4}$ is smaller than $\varepsilon=1.0\times10^{-3}$ of SOT.

By using the inequality (\ref{eq:req}), we derived the tolerances for the physical properties of the instrument.
The tolerance of the waveplate angle is determined as the angle error induced in ${\bf X_{r}}$ at which one of the matrix elements violates inequality (\ref{eq:req}).
In this case, the matrix element representing the cross-talk between Stokes Q and U limits the error to $0^{\circ}.7$.

\subsection{System and component calibration}
In the following sections, we present three experiments, experiment 1 to 3, of TEM's polarization calibration.
Experiments 1 and 3 are end-to-end calibrations and experiment 2 is component calibration. 
\subsubsection{Experiment 1: System calibration}
We have carried out an end-to-end polarization calibration of the instrument twice so far.
We performed the first experiment on 2019 May 23rd and 24th and the second one on 2021 June 10th.
We hereafter call these experiments using the whole instrument as experiment 1.
In 2019, we placed the well-calibrated sheet polarizers (linear, right-hand circular, and left-hand circular) at the entrance of the telescope.
The sheet polarizers used for the test were HN38 (linear), HNCP37R (right-hand circular), and HNCP37L (left-hand circular), all provided by 3M Corporation.
These are the same polarizers used for the polarization calibration of Hinode/SOT \citep{Ichimoto2008}.
The polarizers can be rotated around the optical axis of the telescope by using a stepping motor with a rotation-angle accuracy of $2^{\circ}.8\times10^{-4}$.
For the circular polarizer test, we used 8 different incident Stokes parameters with polarizer angles of $0^{\circ}.00, \pm22^{\circ}.50, \pm45^{\circ}.00, \pm67^{\circ}.50,$ and $90^{\circ}.00$ with respect to the mechanical origin.
For the linear polarizers test, we used 16 different incident Stokes parameters corresponding the polarizer angles of $0^{\circ}.00, \pm11^{\circ}.25, \pm22^{\circ}.50, \pm33^{\circ}.75, \pm45^{\circ}.00, \pm56^{\circ}.25, \pm67^{\circ}.50, \pm78^{\circ}.75$ and $90^{\circ}.00$.
In 2021, the same sheet polarizers (linear and right-hand circular) were also placed at the entrance.
We used 8 different Stokes parameters for both circular and linear polarizers ($0^{\circ}.00, \pm22^{\circ}.50, \pm45^{\circ}.00, \pm67^{\circ}.50,$ and $90^{\circ}.00$ ).
At each rotating position of the sheet-polarizer, we measured Stokes parameters using the normal observation procedure: a waveplate rotation frequency of $1.7$ $\mathrm{Hz}$, a frame rate of $30$ $\mathrm{fps}$, 225 frames at each wavelength, and an exposure time of $4$ $\mathrm{msec}$.
Here we note that in the TEM's observation scheme, we obtain the orientation angle of the retarder's fast axis by using an encoder at every exposure.
The error between exposure and the encoder record timing is about $20$ $\mu$sec, and it corresponds to the angle error of $0^{\circ}.01$ with a rotaion period of $1.7$ $\mathrm{Hz}$.
This is small enough comparing to the rotating angle tolerance of $0^{\circ}.7$.

In order to avoid any significant polarization signal from active regions, the telescope was pointed to the quiet region at the center of the solar disk.
We investigated the polarimeter response matrix at the four wavelengths of $6302.5$ $\mathrm{\AA}$ $\pm0.080$ $\mathrm{\AA}$, $\pm0.160$ $\mathrm{\AA}$ in 2019, and at the six wavelengths of $6302.5$ $\mathrm{\AA}$ $\pm0.050$ $\mathrm{\AA}$, $\pm0.120$ $\mathrm{\AA}$, $\pm0.180$ $\mathrm{\AA}$ in 2021. 

\subsubsection{Experiment 2: Polarizing beam splitter}
\begin{figure}[h]
    \begin{center}
      \includegraphics[bb= 0 0 1250 740, width=120mm]{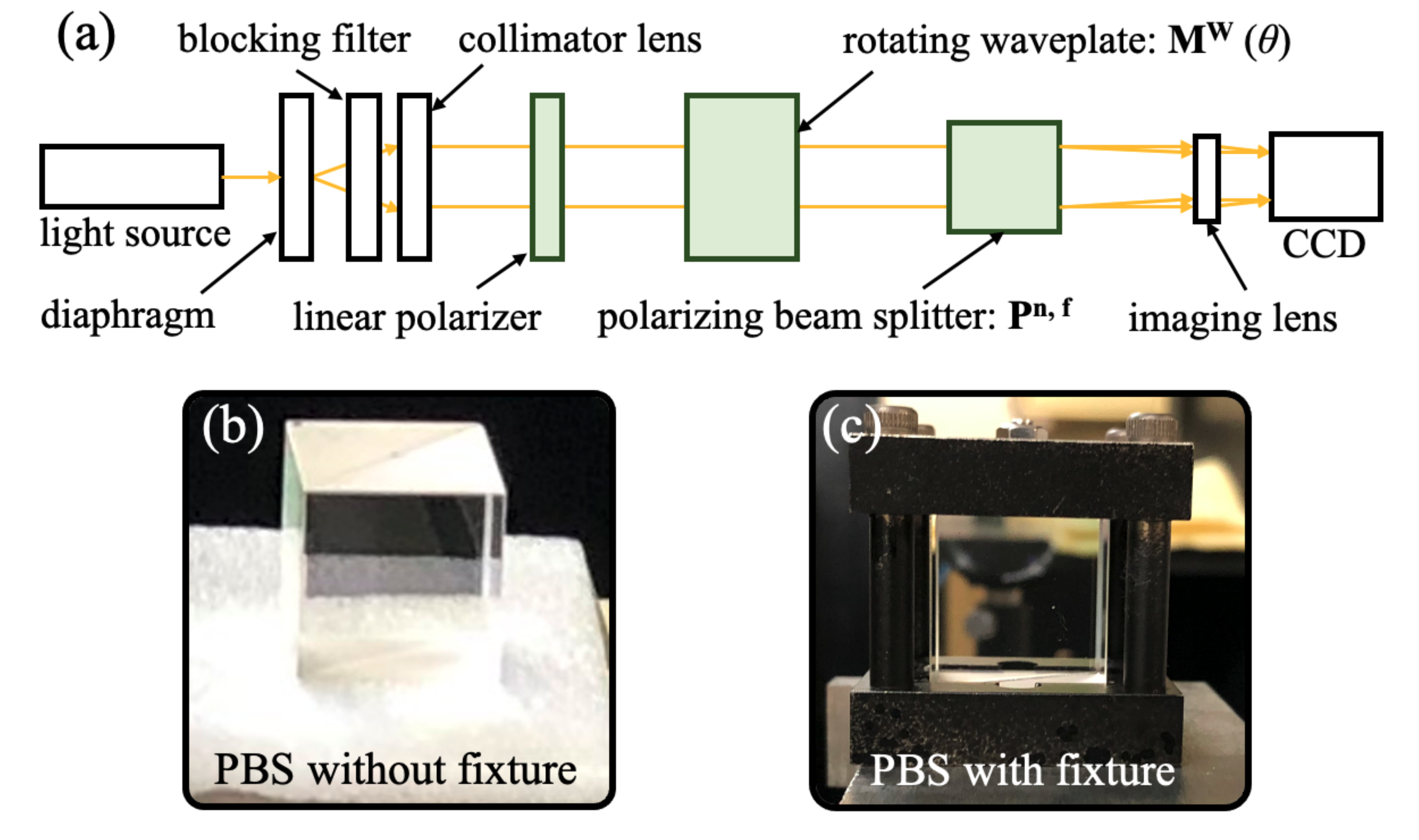}
      \end{center}
    \caption{(a) Experiment setup of the PBS's polarization property evaluation. (b) and (c) show the picture of the PBS without and with fixture, respectively.}\label{fig:exp3}
\end{figure}

We also took the PBS out of the telescope, and examined spatial variations of polarization properties of the PBS in the laboratory.
The optical configuration of the test is shown in figure \ref{fig:exp3} (a).
In this experiment, we used a light source (SIGMA KOKI IMH-250) which is a metal halide lamp with a color temperature of $7500$ $\mathrm{K}$.
We placed a bandpass filter (ANDV9062) which has $3$ $\mathrm{\AA}$ FWHM centered at $6302$ $\mathrm{\AA}$.
The beam was collimated into linearly polarized light by transmission through the collimator lens and the linear polarizer (VLS-200-IR).
After the linear polarizer, a rotating waveplate is placed. 
The intensity of the beam through the PBS was recorded with the same type of CCD camera used in the telescope (Prosilica GE1650), where the surface of the PBS was focused on the camera by an imaging lens.
Moreover, we carried out this experiment with PBS with and without a mount jig to qualitatively understand an effect of the induced mechanical stress of the PBS on its polarization characteristics.

\subsubsection{Experiment 3: Variation in a day}
To evaluate the variation of the response matrix in a day, we performed the end-to-end polarization calibration on 2021 June 9th and 10th.
On June 9th, we placed the linear sheet polarizer (HN38) in front of the telescope and obtained the polarization modulation data from 01:48 to 06:59 UT.
On June 10th, we similarly placed the right-hand sheet polarizer (HNCP37R) and obtained the polarization modulation data from 02:29 to 05:33 UT.
For these measurements, the telescope was again pointed to the disk center and the wavelengths were tuned to $6302.5$ $\mathrm{\AA}$ $\pm0.050$ $\mathrm{\AA}$, $\pm0.120$ $\mathrm{\AA}$, $\pm0.180$ $\mathrm{\AA}$.
The temporal cadence was $\sim200$ $\mathrm{sec}$, and the exposure time of each frame was $4$ $\mathrm{msec}$.

\section{Data analysis and results}
\subsection{Polarimeter response matrix}
In order to calibrate the instrument polarization of the dual-beam system, we have to correct the gain table of the two beams at first. This is done as follows.
In the data analysis of experiment 1, we fitted the observed intensities $I_{\pm}(\theta)$ with a sinusoidal function for both cameras independently.
For example, in the test with an incident Stokes vector of ${\bf S_{in}}=(1,1,0,0)^{\mathrm{T}}$, significant signals in $\cos4\theta$ amplitude corresponding to the input of $Q_{\mathrm{in}}$ are observed by both cameras.
Thus the difference of the spatial distribution of the coefficient of $\cos4\theta$ component between the two cameras can be used to correct the gain table of the two beams.
Similarly, we examined the spatial variation of the most dominant modulation coefficient derived from a sinusoidal fitting for the input Stokes parameters $\pm Q, \pm U,$ and $\pm V$.
We then deduced the gain ratio map of two cameras $g(x,y)$ as follows:
\begin{equation}
  g=\frac{1}{6}\sum_{n=1}^{6}\frac{C_{1,n}}{C_{2,n}}, 
\end{equation}
where $C_{1,n}$ and $C_{2,n}$ represent the most dominant modulation coefficients corresponding to the input Stokes parameters of camera 1 and camera 2, respectively, and $n$ represents the six input Stokes parameters $\pm Q$, $\pm U$, and $\pm V$.

\begin{figure}[h]
    \begin{center}
      \includegraphics[bb= 0 0 1120 400, width=120mm]{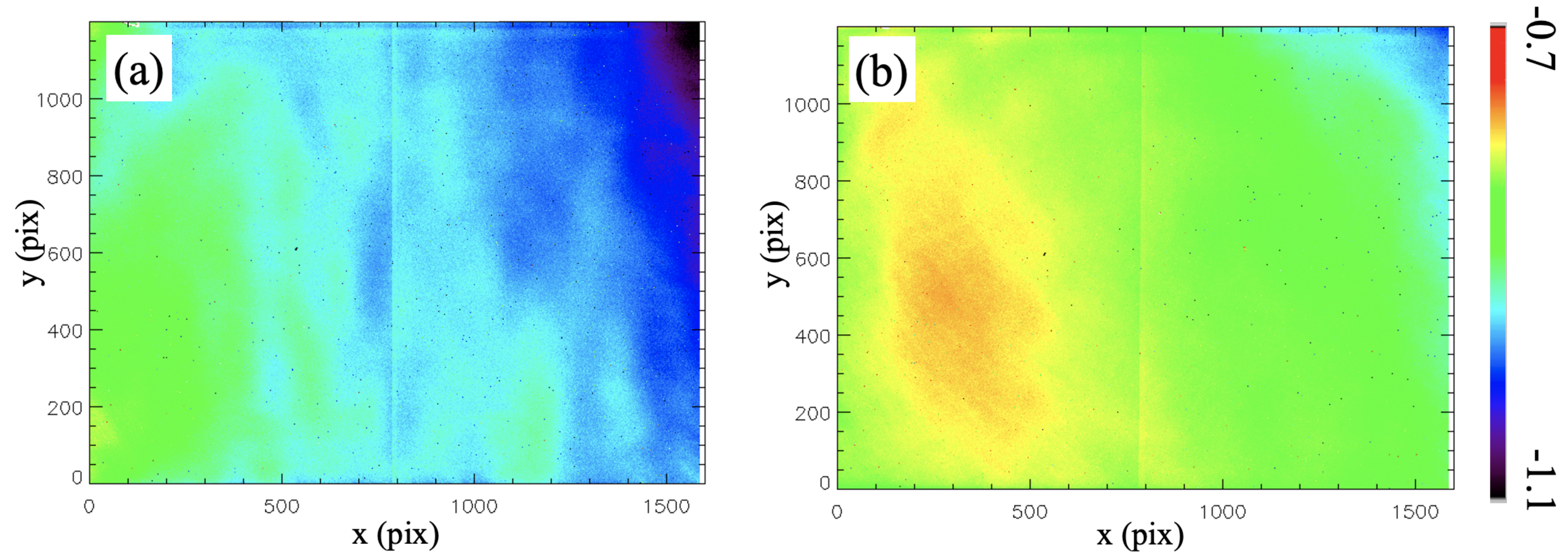}
      \end{center}
    \caption{(a) Gain ratio map through SMART-TEM’s field of view obtained from the experiment in 2019. (b) Gain ratio map obtained from the experiment in 2021.}\label{fig:gain}
\end{figure}

In figure \ref{fig:gain} (a) and (b), we show the gain ratio maps obtained in 2019 and 2021, respectively.
We can clearly see the difference.
In between the two experiments, we took the PBS out of the telescope and set it again in the telescope, as we mentioned above.
Thus, the difference in gain ratio may be due to the change of mechanical stress of the PBS that can change the linear diattenuation of the cube.

We next derived the Stokes vectors (products of the polarimeter, ${\bf S_{out,m}}=(I_{\mathrm{out,m}}, Q_{\mathrm{out,m}}, U_{\mathrm{out,m}}, V_{\mathrm{out,m}})^{\mathrm{T}}$) for each input $m$ with the theoretical model given in equations (\ref{eq:sum}) and (\ref{eq:sub}).
Here we note that $m$ denote the difference of the polarizer angle which we placed in front of the telescope, and $m=1, ..., 32$ for the calibration in 2019 and $m=1,...,16$ for 2021.
The known input Stokes vectors (${\bf S_{in,m}}=(I_{\mathrm{in,m}}, Q_{\mathrm{in,m}}, U_{\mathrm{in,m}}, V_{\mathrm{in,m}})^{\mathrm{T}}$) for each measurement $m$ were simply calculated as follows:

\begin{equation}
    {\bf S_{in,m}} = {\bf M_{pol,m}} I_{\mathrm{out,m}} {\bf E^{I}}, 
\end{equation}
where ${\bf E^{I}}=(1,0,0,0)^{\mathrm{T}}$ and ${\bf M_{\mathrm{pol,m}}}$ correspond to the Mueller matrices of the sheet polarizers.
Those matrices were measured by using the Mueller Matrix Spectro-Polarimeter \citep{Ichimoto2006}.
The spatial uniformity of the polarizers of $\phi=250$ $\mathrm{mm}$ is confirmed by visual inspection by overlapping two polarizers.
The polarizer's axis with respect to the polarization analyzer is self-determined in the analysis of the the calibration data with 16 different polarizer angles.
We determined the $+Q$ axis from the phase of $4\theta$ components.

We finally derived the polarimeter response matrix (${\bf X_{r}}$) by the least-square method with the equation below:

\begin{equation}
    {\bf S_{in,m}} = {\bf X_{r}^{-1}} {\bf S_{out,m}}. \nonumber
\end{equation}

\begin{figure}[h]
    \begin{center}
    \includegraphics[bb= 0 0 1150 450, width=160mm]{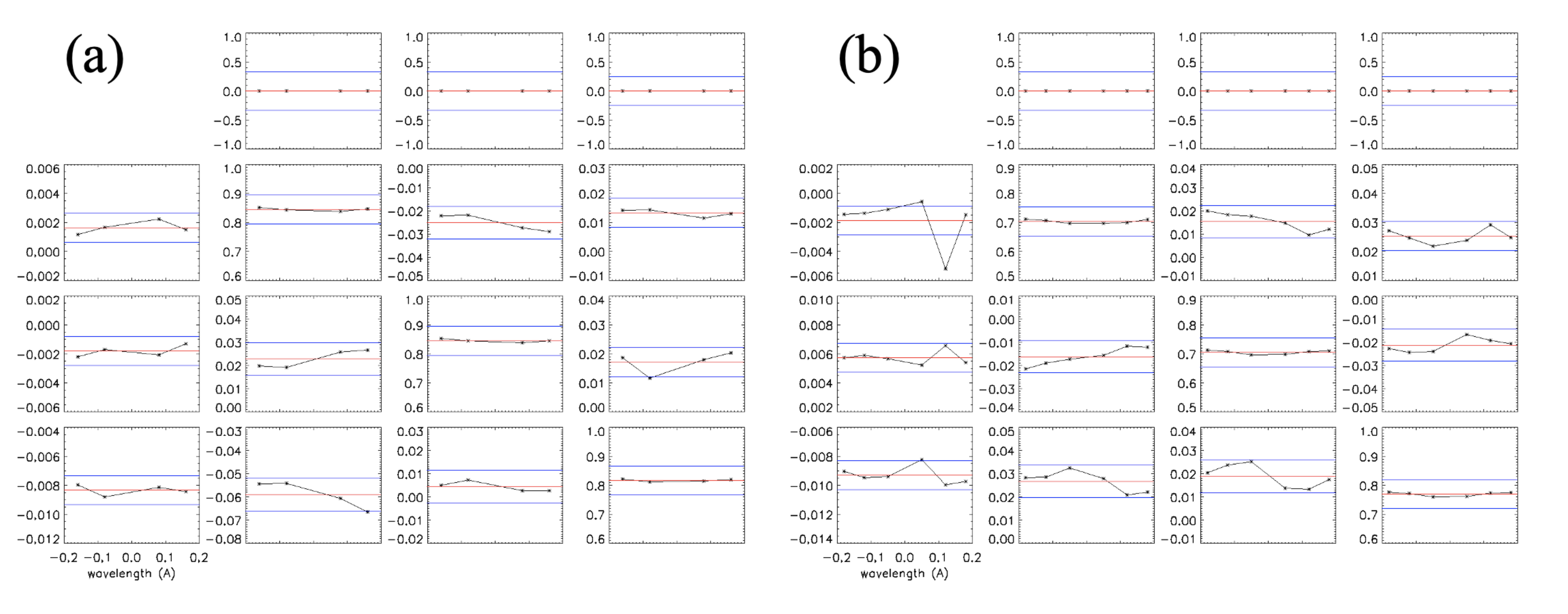}
    \end{center}
    \caption{Wavelength variation of the spatial mean of each polarimeter response matrix component. The line center is $6302.5$ $\mathrm{\AA}$. Red and blue lines correspond to the mean value through four wavelength points and the required accuracy, respectively. (a) Experiment with observation points of four separate wavelengths. (b) Experiment with observation points of six separate wavelengths.}\label{fig:XmatC}
\end{figure}

\begin{figure}[h]
    \begin{center}
    \includegraphics[bb= 0 0 1150 450, width=160mm]{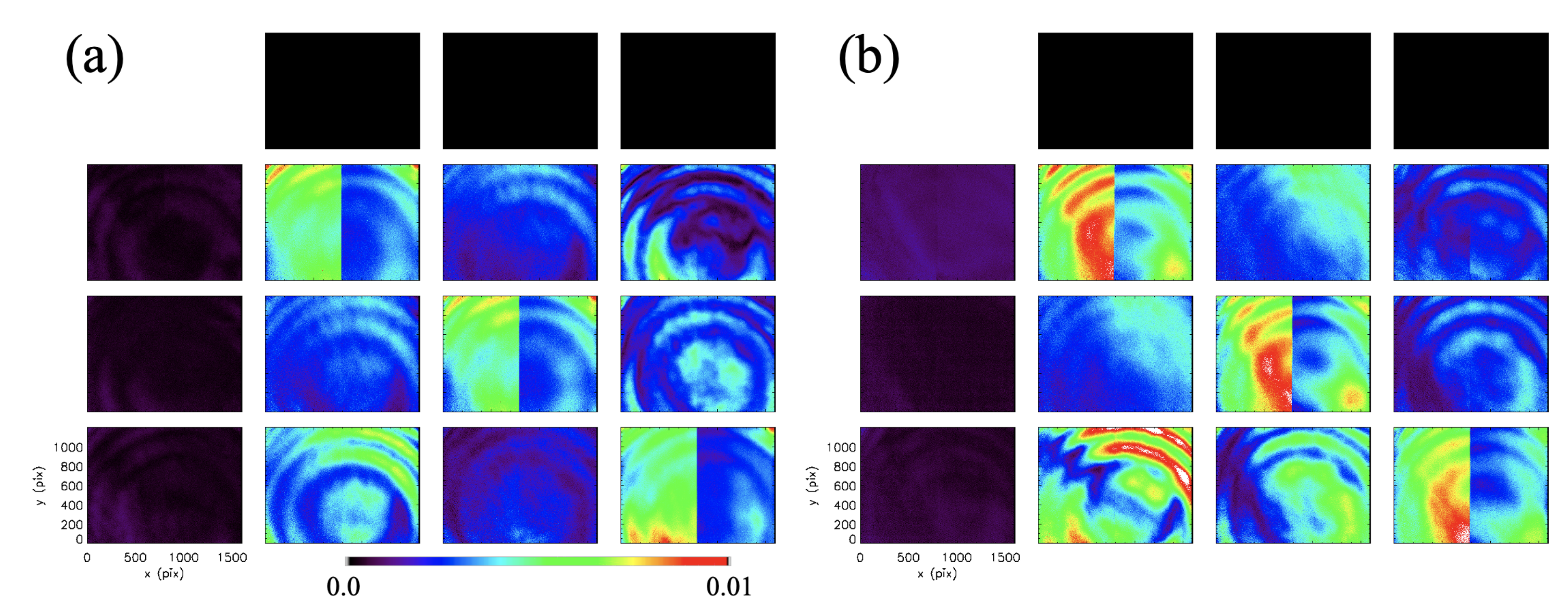}
    \end{center}
    \caption{Standard deviation of polarimeter response matrix over the wavelength points through the TEM's field of view. (a) Experiment in 2019 with observation points of four separate wavelengths. (b) Experiment in 2021 with observation points of six separate wavelengths.}\label{fig:XmatD}
\end{figure}

Figure \ref{fig:XmatC} (a,b) show the spatially averaged polarimeter response matrices as a function of wavelength obtained in 2019 and 2021 respectively.
Each panel corresponds to each component of the polarimeter response matrix.
We obtained the calibration data with four wavelength points of $6302.5\pm 0.080$, and $\pm 0.160$ $\mathrm{\AA}$ in 2019, and with six wavelength points of $6302.5\pm 0.050$, $\pm 0.120$ $\mathrm{\AA}$, and $\pm 0.180$ $\mathrm{\AA}$ in 2021.  
In addition, we examine the wavelength dependence of the polarimetric response matrix across the field of view.
We calculated the standard deviation of each component of the polarimetric response matrices obtained at four or six wavelength points.
The distributions of the standard deviation over the TEM's FOV are shown in figure \ref{fig:XmatD} (a,b).
We can see that the difference through four or six wavelength points is mostly below $0.5\%$.
According to the results, we find that the variations of the polarimeter response matrix components through wavelength are smaller than the required accuracy, which are given in equation (\ref{eq:req}), except the ${\bf X_{12}^{-1}}$ of $6302.8+0.120$ $\mathrm{\AA}$ deduced from calibration in 2021.
In addition, the variations through the wavelength points were smaller than a difference between two calibrations in 2019 and 2021, which can be regarded as measurement error.
Thus, the polarimeter response matrix that averaged through these four or six wavelength points can be a representative of the TEM's response matrix.

\begin{figure}[h]
    \begin{center}
    \includegraphics[bb= 0 0 1150 450, width=150mm]{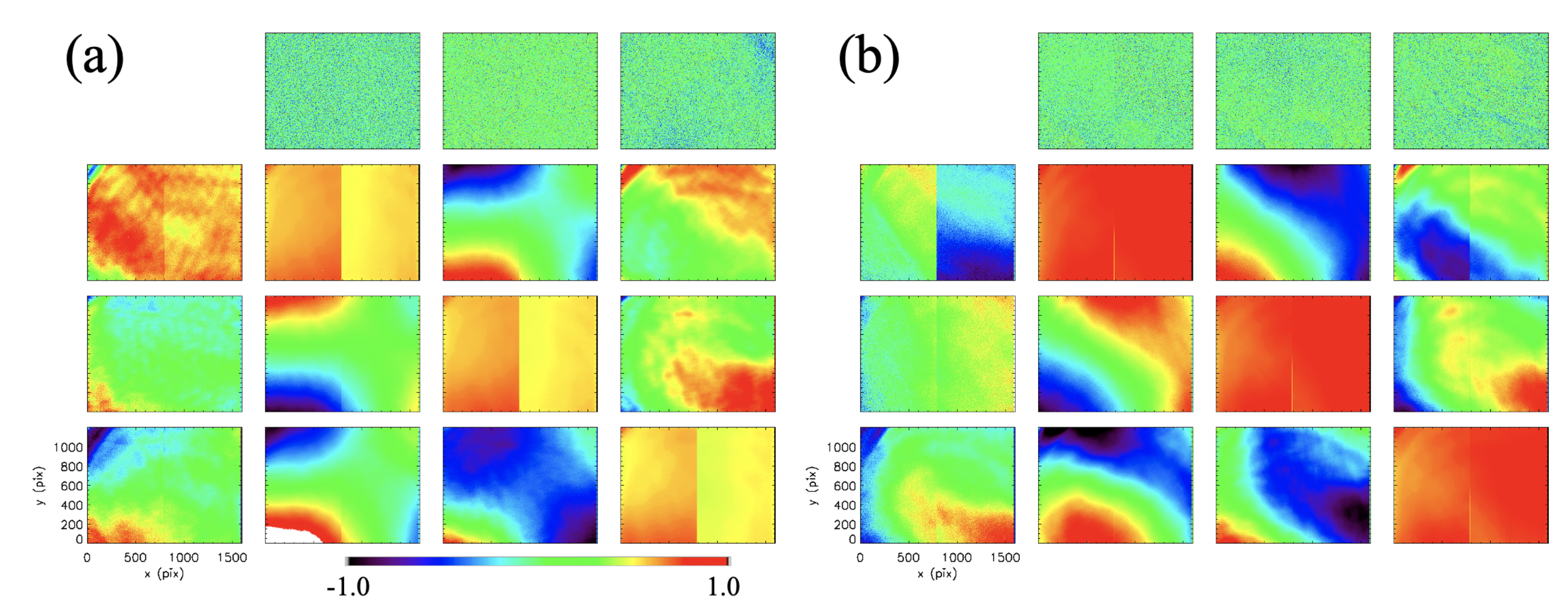}
    \end{center}
    \caption{Polarimeter response matrix ${\bf X_{r}^{-1}}$ map through the TEM's field of view. (a) Experiment in 2019 with observation points of four separate wavelengths. (b) Experiment in 2021 with observation points of six separate wavelengths.}\label{fig:Xmat}
\end{figure}

Figure \ref{fig:Xmat} (a,b) display the polarimeter response matrices averaged over wavelength as a function of field of view.
In the figures, we find significant spatial variations across the field of view.
In order to discuss the spatial distribution quantitatively, we calculate the standard deviation ($\sigma{\bf X_{r,1,2}^{-1}}$) of each component over the FOV for each experiment in 2019 and 2021,
\begin{eqnarray}
    \sigma{\bf X_{r,1}^{-1}} =
    \left(
    \begin{array}{cccc}
        - & 0.000 & 0.000 & 0.000\\
        0.001 & 0.084 & 0.026 & 0.015\\
        0.001 & 0.026 & 0.084 & 0.011\\
        0.002 & 0.067 & 0.008 & 0.082
    \end{array}
    \right), 
\end{eqnarray}
\begin{eqnarray}
    \sigma{\bf X_{r,2}^{-1}} =
    \left(
    \begin{array}{cccc}
        - & 0.000 & 0.000 & 0.000\\
        0.003 & 0.074 & 0.031 & 0.018\\
        0.002 & 0.029 & 0.074 & 0.013\\
        0.004 & 0.076 & 0.027 & 0.080
    \end{array}
    \right). 
\end{eqnarray}
Spatial variations in some components are larger than the requirements given in equation (\ref{eq:req}).  In other words, we can not regard the obtained polarimeter response matrices as uniform across the field of view.
Thus, we need to calibrate the output Stokes vectors pixel by pixel \citep{Yamasaki2019}.  

\begin{figure}[h]
    \begin{center}
    \includegraphics[bb= 0 0 1150 45, width=150mm]{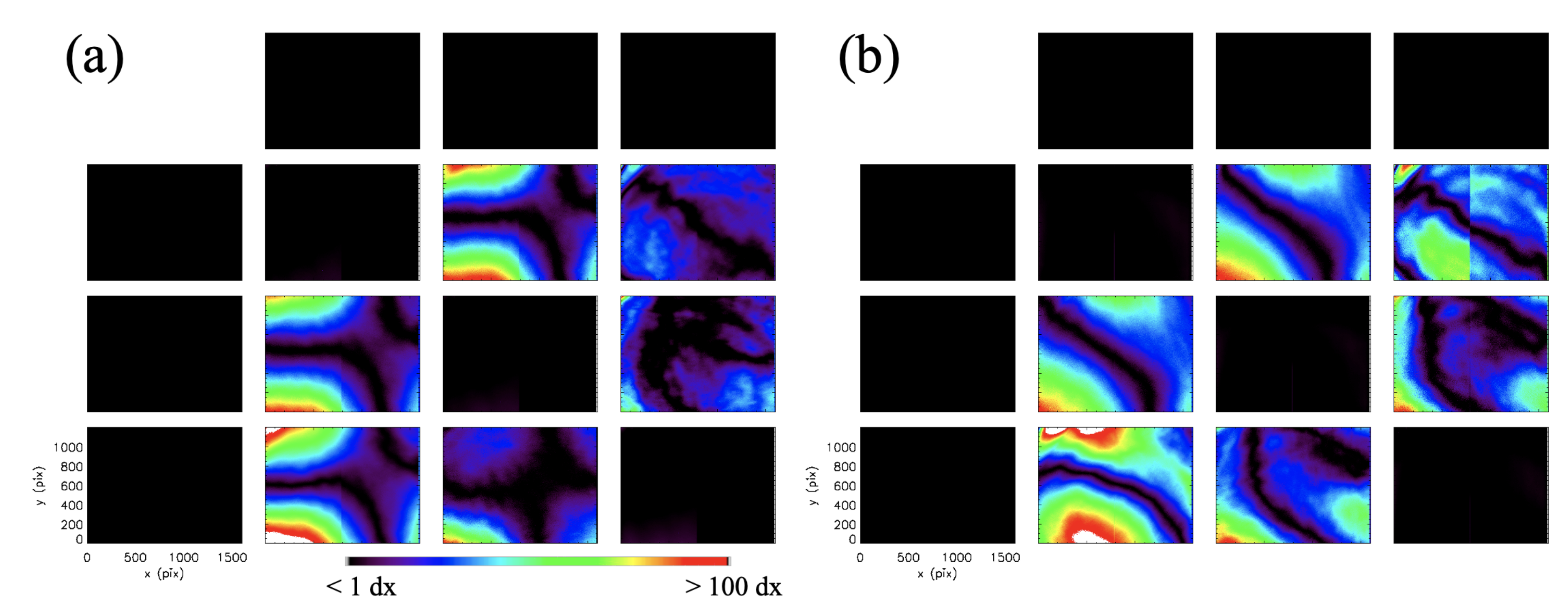}
    \end{center}
    \caption{Deviation from the mean of each component through the TEM's field of view. (a) Experiment in 2019 with observation points of four separate wavelengths. (b) Experiment in 2021 with observation points of six separate wavelengths.}\label{fig:XmatE}
\end{figure}

In order to clearly see the spatial distribution of each component of the response matrices, we show the deviation from the spatially averaged value in figure \ref{fig:XmatE}.
Panels (a) and (b) in the figure \ref{fig:XmatE} correspond to the results of calibration in 2019 and in 2021, respectively. 
In panel (a), we can see that $X^{-1}_{23}$, $X^{-1}_{32}$ and $X^{-1}_{42}$ show similar spatial variation.
We note that those components represent the crosstalk between $Q$ and $U$, and the crosstalk from $Q$ to $V$.
In panel (b), we can see a similar spatial variation in $X^{-1}_{23}$ and $X^{-1}_{32}$, and they resemble the component shown in panel (a).
We also found that some components in figure \ref{fig:Xmat} show a gap at the center of the field of view.
We note that, however, this gap is far below the tolerance as shown in figure \ref{fig:XmatE} and this is not critical on the polarization property of the TEM. 

\subsection{Effect of the mechanical stress of the PBS}
As we mentioned in the previous section, we measured the polarization properties of the PBS by changing the mechanical stress of the PBS in the experiment 2.
If we assume that the spatial variation of the polarization properties of the components other than the PBS is known and small enough to be neglected, then the intensity observed at the CCD camera can be written as follows:
\begin{eqnarray}
  I(\theta)&=&{\bf E^{I}} {\bf P^{f,n}} {\bf M^{W}(\theta)} {\bf S_{in}} \nonumber \\
              &=& I_{\mathrm{in}}+\frac{1+\cos\delta'}{2}Q_{\mathrm{in}}+p_{14}^{\mathrm{f,n}}\cos\delta'V_{\mathrm{in}} \nonumber \\
              &+& (-V_{\mathrm{in}}+p_{14}^{\mathrm{f,n}}Q_{\mathrm{in}})\sin\delta'\times\sin2\theta \nonumber \\
              &-& p_{14}^{\mathrm{f,n}}\sin\delta'U_{\mathrm{in}}\times\cos2\theta \nonumber \\
              &+& \frac{1-\cos\delta'}{2}U_{\mathrm{in}}\times\sin4\theta+\frac{1-\cos\delta'}{2}Q_{\mathrm{in}}\times\cos4\theta, \label{eq:exp3}
\end{eqnarray}
where ${\bf E^{I}}=(1,0,0,0)^{\mathrm{T}}$, ${\bf S^{\mathrm{in}}}=(I_{\mathrm{in}},Q_{\mathrm{in}},U_{\mathrm{in}},V_{\mathrm{in}})^{\mathrm{T}}$ is the incident Stokes vector, ${\bf M^{W}(\theta)}$ is the Mueller matrix of the rotating waveplate, ${\bf P^{f,n}}$ are the Mueller matrices of the PBS with f and n representing the PBS with and without the fixture.
$\delta'$ represents the retardation of the waveplate that we employed in experiment 2, and $\delta'\simeq127^{\circ}$.

\begin{figure}[h]
    \begin{center}
    \includegraphics[bb= 0 0 1100 640, width=120mm]{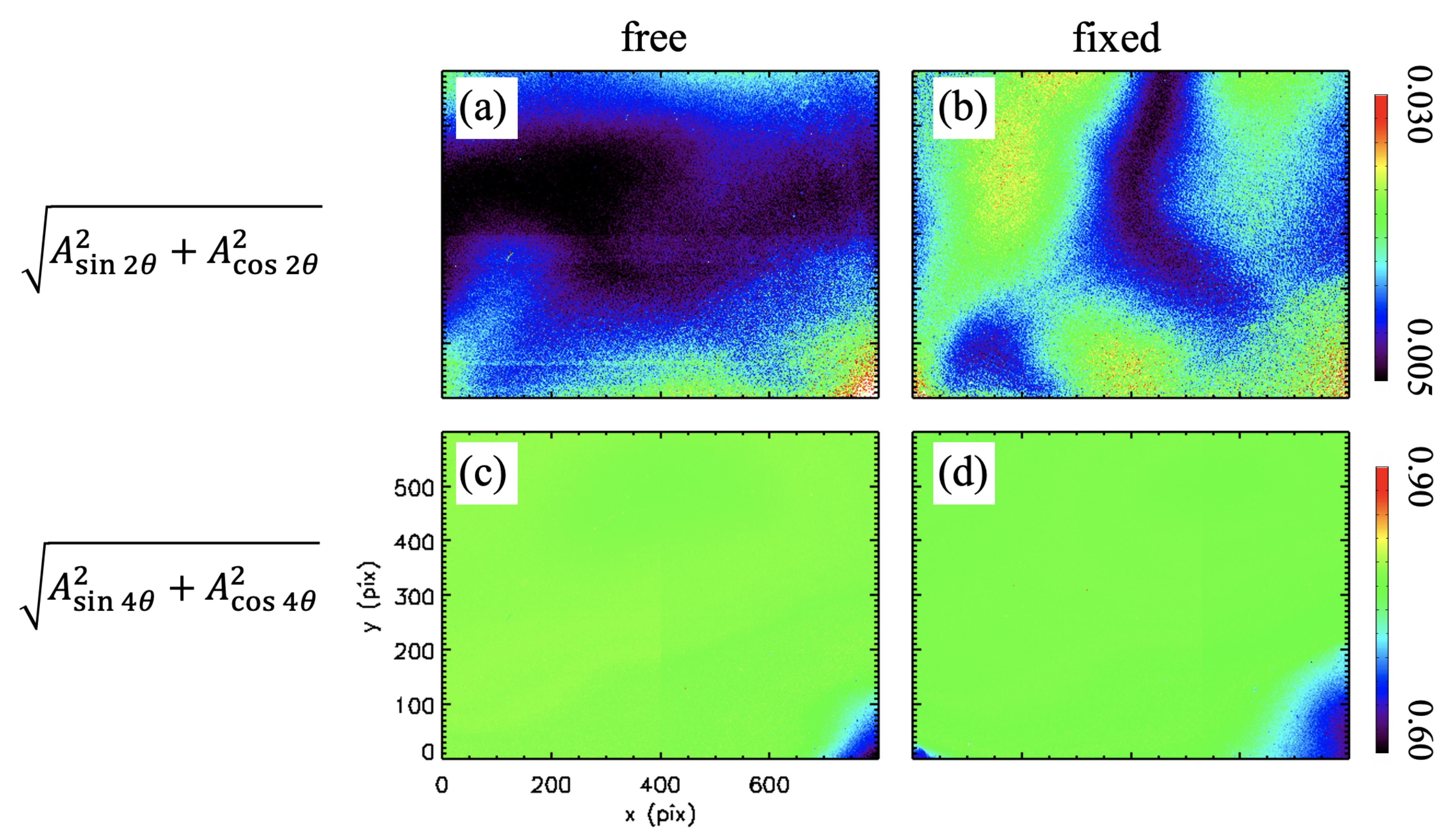}
    \end{center}
    \caption{The sinusoidal fitting amplitude map through the ROI in experiment 2. (a) The absolute value of $2\theta$ component with the free PBS, (b) $2\theta$ component with the fixed PBS, (c) $4\theta$ component with the free PBS, (d) $4\theta$ with the fixed PBS, All are normalized by $\mathrm{const.}$ term of each fitting result.}\label{fig:exp3res}
\end{figure}

The results of the fitting with a sinusoidal function are shown in figure \ref{fig:exp3res}.
Panels (a) and (c) show the spatial distribution of the amplitude of $2\theta$ component ($(A_ {\sin2\theta}^2+A_{\cos2\theta}^2)^{1/2}$) and that of $4\theta$ component ($(A_ {\sin4\theta}^2+A_{\cos4\theta}^2)^{1/2}$), respectively, without fixture (see figure \ref{fig:exp3} (b)).
Panels (b) and (d) show the spatial distribution of the amplitude of $2\theta$ component and that of $4\theta$ component, respectively, with fixture (see figure \ref{fig:exp3} (c)). 
In the figure \ref{fig:exp3res}, we can find that the absolute signal of $4\theta$ is far greater than that of $2\theta$ in both cases of free and fixed PBS.

As for the $2\theta$ component, we find not only a significant signal in $2\theta$ amplitude in each of the measurements, but also a difference in the spatial variation of $2\theta$ amplitude between the two measurements (see figure \ref{fig:exp3res} (a) and (b)).
As shown in equation (\ref{eq:exp3}), the amplitude of $\sin2\theta$ includes not only a $V_{\mathrm{in}}$ signal but also the product of $Q_{\mathrm{in}}$ and $p_{14}^{\mathrm{f,n}}$, the circular diattenuation of the PBS.
In addition, the amplitude of $\cos2\theta$ corresponds to the signal of $U_{\mathrm{in}}$ and it also includes the circular diattenuation of the PBS ($p_{14}^{\mathrm{f,n}}$) (see equation (\ref{eq:exp3})).
Since the amplitude of $2\theta$ component shows $\sim1\%$, the PBS could consist of a $\sim1\%$ circular diattenuation.
Additionally, the significant difference in the spatial pattern found between panels (a) and (b) could be caused by differing stress in the PBS. 
Thus, the circular diattenuation of the PBS could be changed by mechanical stress of the PBS.

As for the $4\theta$ component, we find no significant spatial variation between the two measurements (see figure \ref{fig:exp3res} (c) and (d)).
There is a spatially averaged amplitude of $\sim0.79$ for both cases.
Since we selected linear polarization as the incident signal in this experiment, the amplitude of the $4\theta$ component, which corresponds to the signal of linear polarization, was expected to be $\sim0.80$ $=(1-\cos\delta')/2$ with $\delta'=127^{\circ}$. 
These results suggest that the PBS works as a linear-diattenuator as we expected. 

\subsection{Variation of the polarization property in a day}
\begin{figure}[h]
    \begin{center}
    \includegraphics[bb= 0 0 1200 900, width=150mm]{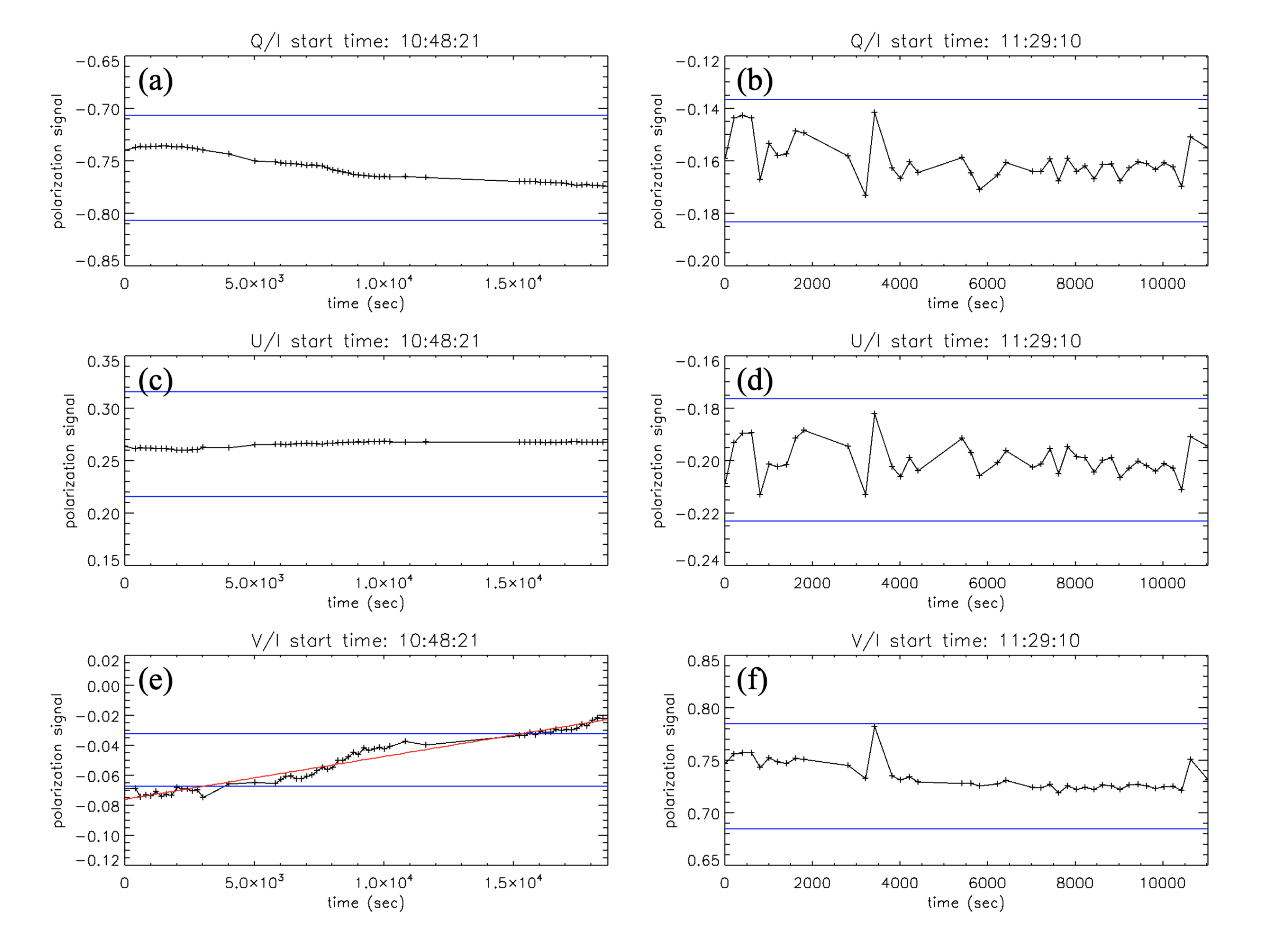}
    \end{center}
    \caption{Temporal evolution of the polarization signals in a day is shown. Each of the blue solid lines show the tolerance corresponding to each polarization signal. (a) $Q_{\mathrm{out}}/I_{\mathrm{out}}$ for linear polarization input. (b) $Q_{\mathrm{out}}/I_{\mathrm{out}}$ for circular polarization input. (c) $U_{\mathrm{out}}/I_{\mathrm{out}}$ for linear polarization input. (d) $U_{\mathrm{out}}/I_{\mathrm{out}}$ for circular polarization input. (e) $V_{\mathrm{out}}/I_{\mathrm{out}}$ for linear polarization input. (f) $V_{\mathrm{out}}/I_{\mathrm{out}}$ for circular polarization input.}\label{fig:dv}
\end{figure}

We deduced the output Stokes vectors from the data obtained in experiment 3.
Figures \ref{fig:dv} (a), (c), and (e) show the daily variations of the Stokes signals of $Q_{\mathrm{out}}/I_{\mathrm{out}}, U_{\mathrm{out}}/I_{\mathrm{out}},$ and $V_{\mathrm{out}}/I_{\mathrm{out}}$, respectively, in the case of linear polarization input.
Figures \ref{fig:dv} (b), (d),  and (f) show the temporal variation of the Stokes signals in the circular polarization input case. 
The signals shown in figure \ref{fig:dv} are averaged over the field of view.  
We find that, in the case of the linear polarization input, the temporal variation of the $V_{\mathrm{out}}/I_{\mathrm{out}}$ shows a variation greater than the tolerance. 
We also find that, for circular polarization, the temporal variation of $Q_{\mathrm{out}}/I_{\mathrm{out}}$ and $U_{\mathrm{out}}/I_{\mathrm{out}}$ show variation less than the tolerance.

\section{Discussion and Conclusion}
In the end-to-end calibration (experiment 1), we obtained the polarimeter response matrices of the TEM in 2019 and in 2021.
We found that the polarimeter response matrices, which are obtained from calibration on different days, have significant spatial variations in their $X_{23}, X_{32},$ and $X_{42}$ components, which correspond to the crosstalk from $Q$ to $U$, from $U$ to $Q$, and from $Q$ to $V$, respectively.
The spatial variations of the components were quite similar to each other.
In the laboratory PBS test (experiment 2), we found that the PBS shows $\sim1$ $\%$ circular diattenuation, and the spatial pattern of circular diattenuation of the PBS is affected by the mechanical stress to the PBS.
Therefore, we suggest that the spatial variation in circular diattenuation of the PBS is due to a combination of  retardation caused by the stress in the cube and linear diattenuation of the dielectric multilayer coating. 

Furthermore, in the end-to-end calibration throughout a day (experiment 3), we found $\sim1\%$ of daily variation in the crosstalk from Stokes $Q,U$ to $V$.
As shown in equation (\ref{eq:sub}), we deduce Stokes $V$ as an amplitude of $\sin2\theta$ of the obtained modulation.
Regarding the theoretical modulation equation, if we consider circular diattenuation of the PBS, the amplitude of $\sin2\theta$ $(A_{\sin2\theta})$ can be written as follows:\\
\begin{eqnarray}
  A_{\sin2\theta} &=& (-V_{\mathrm{in}}+pQ_{\mathrm{in}})\times\sin\delta,  
\end{eqnarray}
where $V_{\mathrm{in}}$ and $Q_{\mathrm{in}}$ are input Stokes signal of $V$ and $Q$, respectively, $p$ corresponds to the circular diattenuation of the PBS and $\delta$ is a retardation of the rotating waveplate. 
This equation shows that one possibile cause of the crosstalk from Stokes $Q,U$ to $V$ is circular diattenuation of the PBS. 
This daily variation result also suggests that the $X_{42}$ and $X_{43}$ components of the polarimeter response matrix were changed on the timescale of a day. 
The temporally linear trend of the polarization degree of $V/I$ is estimated as $\sim 3.0\times10^{-6}$ $\mathrm{sec^{-1}}$ (see red line in figure \ref{fig:dv} (e)).
This means that, in order to meet the required accuracy of $1.8\times10^{-2}$ for the cross talk between Stokes $Q$, $U$, and $V$, we need to obtain calibration data every $\sim100$ $\mathrm{minutes}$.
However, polarization calibration every $100$ $\mathrm{minutes}$ is not a realistic option. 
Thus, we are designing a stress-free mount of the PBS. 

In this paper, we report the results of end-to-end and component polarization calibration of the TEM on SMART at Hida Observatory.
Since the polarization properties of the TEM change both in a short term of a day and in the long term of a year, we suggest that a follow-up investigation of the polarization to meet the polarization accuracy is necessary for our future work. 
The TEM's polarization calibration data presented in this paper are accessible in \url{https://www.hida.kyoto-u.ac.jp/SMART/}.

\begin{ack}
  We thank Mr. Y. Nakatani, Mr. G. Kimura, Ms. K. Hirose, Mr. Y. Kida, and Dr. K. Otsuji for their kind support of the experiments.
  We are grateful to Mr. Cannon Bryce for checking this manuscript.
  This work was supported by the “UCHUGAKU” project of the Unit of Synergetic Studies for Space, Kyoto University.
  This work was also supported by JSPS KAKENHI Grant Number JP21J14036 and JSPS KAKENHI 20684005, Grant-in-Aid for Young Scientist (A).
\end{ack}

\end{document}